\documentclass[a4paper,10pt]{article}
\usepackage[dvips]{graphicx}
\usepackage{braket}
\usepackage{bbm}
\usepackage{amssymb}
\usepackage{amsmath}

\numberwithin{equation}{section}

\newtheorem{theorem}{Theorem}[section]

\newtheorem{lemma}{Lemma}[section]

\newenvironment{proof}
 { \begin{flushleft}
   \textbf{Proof:}
   \begin{list}{}{}%
   \item
 }%
 { \end{list} \vspace{-30pt}
   \begin{flushright} $\blacksquare$ \end{flushright}
   \end{flushleft}
 }

\newenvironment{example}
 { \begin{flushleft}
   \textbf{Example:}
   \begin{list}{}{}%
   \item
 }%
 { \end{list} \vspace{-30pt}
   \begin{flushright} $\diamondsuit$ \end{flushright}
   \end{flushleft}
 }

\DeclareMathOperator{\Tr}{Tr}
 
 \DeclareMathOperator{\RE}{Re}

\newcommand{\elim}[1]{\underset{#1}{\epsilon\textrm{-}\lim}\;\:}

\begin{document}


\thispagestyle{empty}

\begin{center}
\noindent{\Large \bf An extension of the Kac ring model
\normalsize}\\
\vspace{10pt}
{\bf Wojciech De Roeck}\footnote{Aspirant F.W.O. Vlaanderen, U. Antwerpen},
{\bf Tim Jacobs}\footnote{corresponding author, email:
{\tt tim.jacobs@fys.kuleuven.ac.be}}
{\bf Christian Maes}
and {\bf Karel Neto\v cn\'y}\footnote{Instituut voor Theoretische Natuurkunde,
 Rijksuniversiteit Groningen, The Netherlands}\\
{Instituut voor Theoretische Fysica\\ K.U.Leuven, B-3001 Leuven, Belgium.} \vspace{6pt}\\
\today\\
\vspace{7pt} \rule{0.92\textwidth}{0.5pt}
\begin{abstract}\noindent
We introduce a unitary dynamics for quantum spins which is an
extension of a model introduced by Mark Kac to clarify the
phenomenon of relaxation to equilibrium.  When the number of spins
gets very large, the magnetization satisfies an autonomous
equation as function of time with exponentially fast relaxation to
the equilibrium magnetization as determined by the microcanonical
ensemble.  This is proven as a law of large numbers with respect
to a class of initial data.  The corresponding Gibbs-von Neumann
entropy is also computed and its monotonicity in time discussed.
\end{abstract}
\rule{0.92\textwidth}{0.5pt}
\end{center}

\section{Relaxation to equilibrium}
\label{sec:introduction}

The Kac ring model was introduced by Mark Kac to clarify how
manifestly irreversible behaviour can be obtained from an
underlying reversible dynamics, \cite{Bricmont,Kac,Thompson}. It
explains via a simple model some of the conceptual subtleties in
the problem of relaxation to equilibrium as for example are
present in the derivation and the status of the Boltzmann equation
for dilute gases. In particular, the Kac dynamics shares some
basic features with a Hamiltonian time-evolution like being
deterministic and dynamically reversible.  In the present paper we
extend that dynamics to a unitary evolution on a finite quantum
spin system. Again, the dynamics remains far from realistic but it
allows a precise formulation and discussion of some features of
relaxation to equilibrium for a quantum dynamics. That is
especially useful and relevant as, in the quantum domain, the
problem of relaxation is beset with even greater conceptual
difficulties. In our framework, relaxation to equilibrium becomes
visible if one can select a small number of macroscopic variables
that typically evolve via autonomous deterministic equations to
take on values that correspond to equilibrium. Typical refers to a
law of large numbers with respect to the initial data. Paradoxes
are avoided by taking serious the fact that relaxation is a
macroscopic phenomenon, involving a huge amount of degrees of
freedom whose evolution is monitored over a realistic time-span.
One should also keep in mind that relaxation to equilibrium goes
beyond questions of return to equilibrium, see \cite{rob}, which
are mostly related to stability of equilibrium states.  A more
general introduction to that and various related problems can be
found in the recent \cite{sew}.\\

In section
\ref{sec:model} we introduce the model and we state the basic
result. Section \ref{sec:discussion} is devoted to a discussion of
related issues. The proofs are postponed to the final section
\ref{sec:proofs}.

\section{Model and results}
\label{sec:model}

\subsection{The model}

Consider $N$ sites on a ring (periodic boundary conditions).
Between any two neighboring sites there is a fixed {\it
scattering} mechanism to be specified below, and at each site, we
find a spin $1/2$ particle.  Time is discrete and at each step the
ring rotates in a fixed direction over one ring segment. Depending
on the segment that each spin crosses, it is scattered to another
state.

For the Hilbert space $\mathcal{H}_N$ we take the $N$-fold product of
copies of $\mathbb{C}^2$, the state space at each site:
\begin{displaymath}
\mathcal{H}_N \equiv \bigotimes_{j=1}^N \mathbbm{C}_j^2
\end{displaymath}
with the standard inner product that defines the Hermitian
conjugate for matrices denoted by the superscript $^\star$. We
write the elements of $\mathcal{H}_N$ as vectors in
$\mathbbm{C}^{2N}$, denoted by $\eta$ or $\eta'$ with components
$\eta_j$ indicating the state of the spin at site $j$.

\subsubsection{Dynamics and observables}
We fix an arbitrary Hermitian matrix $H\in \mathbbm{C}^{2\times 2}$,
called single site Hamiltonian and we construct the unitary matrix
\begin{equation}
\label{eq:QKM-UnitaryOneSite}
U \equiv e^{iH}, \;\; U^\star = e^{-iH}
\end{equation}
To each segment on the ring connecting the two nearest neighbors
$(j,j+1)$ there is associated a scattering variable $ \epsilon_j
\in \{0,1\}$.  The configuration of scatterers is denoted by
$\epsilon \equiv (\epsilon_1,\ldots,\epsilon_N)$ and does not
change in time. We define the unitary matrix $U_j^{\epsilon_j}$ on
$\mathbbm{C}_j^2$ as
\begin{displaymath}
U_j^{\epsilon_j} \equiv \left\{ \begin{aligned}
& \mathbbm{1} && \qquad \qquad \textrm{for } \epsilon_j = 0 \\
& U && \qquad \qquad \textrm{for } \epsilon_j = 1
\end{aligned} \right.
\end{displaymath}
The superscript $\epsilon_j$ can thus be read as a power and
$U_j^{\epsilon_j} = e^{i \epsilon_j H_j}$ with $H_j$ a copy of $H$
working on $\mathbbm{C}_j^2$.  This defines the scattering
mechanism.\\

The rotation of the ring is implemented by the operation $R$ on
$\mathcal{H}_N$ which transforms every vector
$\eta =(\eta_1,\eta_2,\ldots,\eta_N)$ into
\begin{displaymath}
R\eta \equiv (\eta_N,\eta_1,\ldots,\eta_{N-1}) \in \mathcal{H}_N
\end{displaymath}
and $R^\star = R^{-1}$, rotation in the opposite direction.
Combining that with the scattering mechanism finally gives rise to
the unitary operator $U_N\equiv U_N(\epsilon)$ on $ \mathcal{H}_N$
via
\begin{displaymath}
U_N \equiv R \, \bigotimes_{j=1}^N U_j^{\epsilon_j}
\end{displaymath}
To be specific we look at positive times and the state at time
$t=1,2,\ldots$ is then obtained from the state $\eta_0
\in\mathcal{H}_N$ at time $t=0$ from
\begin{displaymath}
\eta_t \equiv U_N(t)\, \eta_0 = U_N(t-1) U_N \, \eta_0
\end{displaymath}
This unitary dynamics on the finite-dimensional $\mathcal{H}_N$ is
rather simple. For many observables (like the total magnetization)
the dynamics is entirely equivalent with leaving the spins in
place and rotating instead the scatterers.  We therefore work with
the more convenient
\begin{equation}
\label{eq:QKM-UnitarySystem}
U_N(t) = \bigotimes_{j=1}^N U_j^{k_j(\epsilon,t)}
\end{equation}
with $k_j(\epsilon,t) \equiv \sum_{n=1}^t \epsilon_{j-n}$ (modulo
$N$). The updating of the spins is independent modulo the fact
that they may have a scattering mechanism in common.\\

The Hamiltonian is a sum of single site contributions,
\begin{equation}\label{hami} H_N \equiv \sum_{j=1}^N H_j
\end{equation}
with $H_j \equiv \mathbbm{1} \otimes \ldots \otimes H \otimes
\ldots \otimes \mathbbm{1}$ (on the $j$-th site) and for
concreteness, we decompose the single-site Hamiltonian in the
Pauli basis
\begin{equation}
\label{eq:QK12_Hamiltonian} H = h_1 \, \sigma_x + h_2 \, \sigma_y
+ h_3 \, \sigma_z = \vec{h} \cdot \vec{\sigma} \qquad \qquad 0\neq
\vec{h} \in \mathbbm{R}^3
\end{equation}
The eigenvalues are $e_{+} = \| \vec{h} \| \equiv h$ and $e_{-} =
-e_{+}$. (We can of course ignore  adding a constant to this
Hamiltonian.)
 Our observables are also built up by adding one-site contributions.
We are given a Hermitian matrix $A \in \mathbbm{C}^{2 \times 2}$
and we construct the system operators
\begin{equation}
\label{eq:sma} A_N \equiv \sum_{j=1}^N \mathbbm{1} \otimes \ldots
\otimes A \otimes \ldots \otimes \mathbbm{1}
\end{equation}
where the $A$ is at the $j$-th position. We will consider the
magnetization vector $\vec{M}_N \equiv (M_N^x,M_N^y,M_N^z)$
defined through the one-site observables $M^{\alpha} =
\sigma_{\alpha}$, the Pauli matrices, with $\alpha \in \{ x,y,z\}$.\\
The restriction of looking only at one-site observables will be
discussed in Section \ref{subsec:norelax}.

\subsubsection{Equilibrium}
The dynamics depends non-trivially on the scatterers but the
equilibrium properties are independent of them.  That is similar
to the situation in the Boltzmann-Grad limit where the hard core
matters dynamically but does not enter in the computations of
energy or pressure.\\
 Energy is conserved in the sense that the
Hamiltonian
commutes with the time evolution: $[U_N(t),H_N] = 0$.\\

Equilibrium is characterized by the microcanonical distribution. Let
$\psi_+, \psi_-$ be the two eigenvectors of $H$ with eigenvalues
$e_+ \geqslant e_-$ and spectral projectors $P_{e_+}$ and $P_{e_-}$
respectively. For arbitrary $e \in [e_-,e_+]$  we select an energy space by
\begin{equation}\label{pe}
P_{e}^N \equiv \frac 1{Z_{e}^N}\,\sum_{e} P_{e_1}
\otimes \ldots \otimes P_{e_N}
\end{equation}
where, in $\sum_e$, we sum over all $(e_j)$ with fixed $\sum_j
e_j$ satisfying $Ne \leqslant \sum_j e_j < e_+ - e_- + Ne$; the
normalization $Z_e^N$ ensures that $\Tr[P_{e}^N]=1$.
Alternatively, we could have summed in \eqref{pe} over all $(e_j)$
with $\sum_j e_j/N$ in a certain interval around $e$ and at the
very end let the interval shrink to zero. The average
\begin{equation}
\label{eq:fmicro}
\braket{A_N}^N_{e} \equiv \mbox{Tr}[ P_{e}^N \, A_N]
\end{equation}
for a Hermitian matrix $A_N$ on $\mathcal{H}_N$ defines the
(finite volume) microcanonical ensemble.\\

The (infinite volume) equilibrium magnetization corresponding to
an energy $e$ is then defined from \eqref{eq:fmicro} to be the
limit
\begin{equation}
\label{eq:micro} \vec{m}_e \equiv  \lim_{N\uparrow +\infty}
\frac1{N} \braket{\vec{M}_N}^N_e
\end{equation}
This limit exists and can easily be computed: it equals
\begin{equation}
\label{eq:cmicro} \vec{m}_e = e \,\frac{\vec{h}}{h^2}
\end{equation}

\subsubsection{Initial data}\label{ind}
Initial data are determined by a density matrix $\rho^N_0$ on
$\mathcal{H}_N$ and by the choice of the scatterers $\epsilon$.
Concerning the state of the spins, let us keep in mind an initial
preparation with a particular magnetization in the $z$-direction,
not in equilibrium. Let $Q_+$ and $Q_-$ be the projectors on spin
up and spin down respectively, with $\sigma_z = Q_+ - Q_-$, to
define
\begin{equation}\label{qm}
Q^N_{m}
\equiv
\sum_{m} Q_{m_1} \otimes \ldots \otimes Q_{m_N}
\qquad \qquad
m \in \{ -1, +1 \}
\end{equation}
where, as in \eqref{pe}, in $\sum_m$, we sum over all $(m_j)$ with
fixed $\sum_j m_j$ satisfying $Nm \leqslant \sum_j m_j < 2 + Nm$.
Equation \eqref{qm} gives the projector on the magnetization space with
magnetization converging to $m$ as $N\uparrow +\infty$.  A
possible initial density matrix is then ($0 < d(m,N) < +\infty$ is
a normalization)
\begin{equation}
\label{eq:ainidens}
\rho^N_0 = \frac 1{d(m,N)}\,Q^N_{m}
\end{equation}
Another (but thermodynamically equivalent) choice would be the
grand-canonical
\begin{equation}\label{grand}
\rho^N_0 = \frac 1{Z_N(\lambda)}\,\exp \big( {\lambda M^z_{N}}
\big)
\end{equation}
and similarly for the other directions of the magnetization. A
more general class of initial density matrices $\rho_0^N$ will be
introduced at the beginning of section \ref{results}. Most
important is that they satisfy a law of large numbers for
observables of the form \eqref{eq:sma} and that the single site
marginal remains well-defined in the
thermodynamic limit, see \eqref{eq:reduc}.\\

Now the dynamics starts and the density matrix at time $t$ is
defined from \eqref{eq:QKM-UnitarySystem}:
\begin{equation}
\label{eq:tdens}
\rho^N_t \equiv \rho^N_t(\epsilon) \equiv  U_N(t) \rho^N_0 U_N(t)^\star
\end{equation}
and now also depends on the scatterers $\epsilon_j$.

We are interested in the magnetization in $\rho_t^N$ and in its
limiting behavior as (first) $N\uparrow +\infty$ and (then) $t
\uparrow +\infty$. In order to take these limits, we also need to
specify the initial (and unchanging) condition of the scatterers.
For this we introduce the set
\begin{equation}
\label{eq:epsset} \Omega_{\mu}^N \equiv \Bigg\{ \epsilon \in
\{0,1\}^N \Bigg| \sum_{j=1}^{N} \epsilon_j = \lceil \mu N \rceil
\Bigg\}
\end{equation}
with  $\lceil \mu N \rceil$ the smallest integer not smaller than
$\mu N$.  The constraint fixes the fraction of {\it active}
scatterers to be about $\mu$.  The probability to select one
particular $\epsilon$ in $\Omega_{\mu}^N$ is uniform:
\begin{equation}
\label{eq:QKM-ProbCanonical} \mathbbm{P}^N_{\mu} \big( \, \epsilon
\,\big) \equiv \frac{1}{ \binom{N}{\lceil N\mu \rceil}}
\end{equation}
All limits will involve that probability distribution.  We say
that a sequence of functions $G_N$ on $\Omega_\mu^N$ typically
takes the value $g_\mu$, written
\begin{equation}\label{typ}
\elim{N\uparrow +\infty} G_N = g_\mu \qquad \textrm{iff} \qquad
\lim_{N\uparrow +\infty} \mathbbm{E}^N_\mu[(G_N-g_\mu)^p] =0,\quad
p=1,2
\end{equation}
where the expectation refers to the probabilities
\eqref{eq:QKM-ProbCanonical}. Via the Chebyshev inequality one can
reformulate \eqref{typ} as a statement about the fraction of
scatterers in $\Omega^N_\mu$ for which $G_N \simeq g_\mu$.  The
last $\simeq$ is made stronger if higher $p$'s in \eqref{typ} are
obtained.  For simplicity, we restrict us here to the average
($p=1$) and to the variance ($p=2$).

\subsubsection{Entropy}
One expects for the given model that the one-particle distribution
satisfies an autonomous equation.  That information is encoded in
the single site density matrix.  For finite $N$, the marginals are
defined through
\begin{equation}\label{eq:marginal}
 \nu^N \equiv \frac{1}{N} \sum_{j=1}^N \Tr_j
\big[ \rho^N \big]
\end{equation}
with $\Tr_j$ the reduced density matrix at site $j$. The Gibbs-von
Neumann  entropy is defined following the ideology of
\cite{QuantumEntropy} through the variational principle
\begin{equation}\label{supp}
S_N(\rho^N) \equiv \sup_{\rho'^N} - \Tr\big[ \rho'^N \log \rho'^N
\big]
\end{equation}
where the supremum is over all density matrices $\rho'^N$ with
marginal $\nu^N$ from \eqref{eq:marginal}. Obviously, the supremum
in \eqref{supp} is attained for the product state $\rho'^N =
\bigotimes \nu^N$ and therefore through \eqref{eq:marginal}, the
quantum entropy equals
\begin{equation}
\label{eq:QGE} S_N(\rho^N) = -N \Tr\Big[ \nu^N \log \nu^N \Big]
\end{equation}
We insert $\rho_t^N$ from \eqref{eq:tdens} in \eqref{eq:QGE} and
put, hoping all goes well,
\begin{equation}\label{entdens}
s(t) \equiv \elim{N\uparrow +\infty} \frac 1{N}  S_N(\rho^N_t)
\end{equation}
as the time-dependent entropy (density). The more frequently
considered von Neumann entropy -Tr $[\rho_t^N \log \rho_t^N]$ does
not change in time and is therefore here less relevant.

\subsection{Results}\label{results}
The conditions for our main result on the dynamics
 \eqref{eq:tdens} are on the level of initial
data, see section \ref{ind}.   We ask that $\rho_{0}^N$ is a
density-matrix that is dispersionfree in the sense that
 for all observables \eqref{eq:sma} and for all continuous $f$
\begin{equation}
\label{eq:initialansatz}
\lim_{N \uparrow +\infty}
\Bigg| \Tr \Bigg[ f \bigg( \frac{A_N}{N} \bigg) \rho_{0}^N  \Bigg]
- f \Bigg( \Tr\bigg[ \frac{A_N}{N} \rho_{0}^N \bigg] \Bigg)\Bigg| = 0
\end{equation}
Furthermore,  we  need the existence of the thermodynamic limit of
the system-averaged one-site marginal defined in
\eqref{eq:marginal}:
\begin{equation}
\label{eq:reduc} \nu_0 \equiv \lim_{N \uparrow +\infty} \frac
1{N}\sum_{j=1}^N \Tr_j \big[ \rho^N_0 \big] \equiv
\frac{\mathbbm{1} + \vec{m}_0\cdot \vec{\sigma}}{2}
\end{equation}
and similarly, we also define the time-evolved version of $\nu_0$,
i.e.,
\begin{equation}
\label{eq:reduct}
 \nu_t \equiv \elim{N \uparrow +\infty} \frac
1{N}\sum_{j=1}^N \Tr_j \big[ \rho^N_t \big]
\end{equation}
 Finally, the sequence $\rho^N_0$ must satisfy a
technical condition.  We write
\begin{equation}
\label{eq:density} \rho_0^{N} = \sum_{w=1}^{r(N)}
\bigotimes_{j=1}^{N} \chi_{j,w}^N
\end{equation}
the initial density matrix as a sum of products of $2\times 2$
matrices and we assume that
\begin{equation}
\label{eq:statebound} \sup_{N \in \mathbbm{N}} \sup_{w=1,..,r(N)
\atop j=1,..,N} \| \chi_{j,w}^N \| \equiv C < \infty
\end{equation}
These three conditions are obviously satisfied by \eqref{grand}.
That \eqref{eq:initialansatz} is also satisfied for
\eqref{eq:ainidens} needs an argument given in the example after
the proof of Lemma
\ref{lemma:large deviations}.\\

The initial magnetization is denoted by $\vec{m}_0 \equiv  \Tr
[\vec{\sigma} \nu_0]$ and the equilibrium magnetization
$\vec{m}_e$ is defined in \eqref{eq:micro}. We denote components
of the magnetization as $m^{\alpha}$, $\alpha \in \{ x,y,z \}$.
The initial energy $e$ is found from $e\equiv \Tr[ H \nu_0]$.

\begin{theorem}~\\
\label{result:relaxation} \!\!\!\!
For all continuous functions $f$ and for initial $\rho^N_0$ satisfying the above
conditions,
\begin{equation}
\label{eq:rel} \elim{N \uparrow \infty} \Tr\Bigg[
f\bigg(\frac{M^{\alpha}_N}{N} \bigg)\,\rho^N_t \Bigg] = f \big(
m^{\alpha}_t \big)
\end{equation}
with
\begin{equation}\label{mrel}
\vec{m}_t \equiv \vec{m}_e + \RE\Bigg[ \bigg(\vec{m}_0 - \vec{m}_e
+ \frac{i(\vec{h}\times\vec{m}_0)}{h} \bigg) \Big(1-\mu+\mu
\exp[i(e_+ - e_-)]\Big)^t\Bigg]
\end{equation}
In particular for $f(x)=x$, the magnetization
\begin{equation}
\label{eq:QK12_Relaxation} \lim_{t \uparrow +\infty} \elim{N
\uparrow +\infty}\frac {1}{N} \Tr\big[\vec{M}_N \, \rho^N_t \big]
= \vec{m}_e
\end{equation}
converges exponentially fast to its equilibrium value, for all
$\mu \in (0,1)$ when $e_+ - e_-$ is not a multiple of $2\pi$.\\
~\\
The entropy \eqref{entdens} equals
\begin{equation}
\label{result:entropy} s(t) = - \frac{1+\|\vec{m}_t\|}{2}
\log\bigg( \frac{1+\|\vec{m}_t\|}{2} \bigg) -
\frac{1-\|\vec{m}_t\|}{2} \log \bigg( \frac{1-\|\vec{m}_t\|}{2}
\bigg)
\end{equation}
where $\vec{m}_t$ is given by \eqref{mrel}.  $s(t)$ is increasing
in time ($H-$theorem) and can be written as
\begin{equation}
\label{eq:QuaEntDensity}
s(t) = \mathcal{S}[\nu_t],\quad
\mathcal{S}[\nu]
 \equiv - \Tr[ \nu \log \nu ]
\end{equation}

\end{theorem}

\section{Discussion}
\label{sec:discussion}

\subsection{Kac ring model}
The Kac ring model in \cite{Kac}, p.99, can be recovered by the
appropriate choice of the Hamiltonian \eqref{eq:QK12_Hamiltonian},
by only looking at the magnetization in the $z$-direction.
Identifying $\eta_i=1$ or $\ket{\uparrow}$ with a white ball and
the state $\eta_i=-1$ or $\ket{\downarrow}$ with a black ball in
the original ring model, the switching between a white and a black
ball corresponds to a spin flip. This is accomplished by the
single-site Hamiltonian (written out in the $\sigma_z$ base):
\begin{displaymath}
H = \frac{\pi}{2} \begin{pmatrix}1 & -1 \\ -1 & 1 \end{pmatrix}
\qquad \Rightarrow \qquad U = \begin{pmatrix} 0 & 1 \\ 1 & 0
\end{pmatrix}
\end{displaymath}
which, up to an irrelevant constant, corresponds to the choice
$\vec{h}=-(\frac{\pi}{2},0,0)$ in \eqref{eq:QK12_Hamiltonian}. The
eigenvalues of this Hamiltonian satisfy $e_{+} - e_{-} = \pi$ and
the equilibrium magnetization is $m^{z}_e=0$.  We start from
randomly sampling the balls (or spins) with a fixed overall color
(or magnetization) in \eqref{eq:ainidens} so that $m^{x}_t =
m^{y}_t = 0$ for all $t$.   Substitution in \eqref{mrel} yields:
\begin{displaymath}
m^z_t = m^z_0 (1- 2\mu)^t
\end{displaymath}

That coincides with the relaxation formula in the original Kac
model. It is interesting to observe that the relaxation to
equilibrium gets slower for all the extensions that we have
considered here.

\subsection{Molecular chaos}\label{moc}

The Sto\ss\-zahlansatz, or \emph{repeated randomization}
hypothesis consists in replacing the real dynamics by an effective
dynamics in which the scattering mechanism is not kept fixed but
gets replaced with an average. In that way, memory is being erased
of where the scatterers are.   An effective dynamics then works on
the single particle level and since the fraction of active
scatterers equals $\mu$, the one-site density matrix $\nu_t$ is
either copied with probability $(1-\mu)$ or is replaced with a
scattered density matrix with probability $\mu$. More precisely,
\begin{equation}
\label{eq:QKM-SZA_CPMap} \Gamma^{\star}(\nu) \equiv ( 1 - \mu )
\nu + \mu \, U \nu \, U^{\star}
\end{equation}
defines a completely positive map (the quantum equivalent of a
discrete time stochastic dynamics, see \cite{AF}).
 The unitary $U$
should be substituted from \eqref{eq:QKM-UnitaryOneSite}.  On the
level of the full density matrix $\rho^N$, we just make the
product of copies of $\Gamma^{\star}$ over all $N$ sites so that a
product state is mapped into a product state: If $\nu^N(j) \equiv
\Tr_j \big[ \rho^N \big]$, then
\begin{equation}
\Lambda^{\star}(\rho^N) \equiv \bigotimes_{j=1}^{N}
\Gamma^{\star}\Big(\nu^N(j) \Big)
\end{equation}
That is the dual of a map $\Lambda$ and we write $\Lambda_t$ for
the map $\Lambda$ applied $t$ times. The effective dynamics thus
amounts to replacing the time evolution $U_N(t)\cdot U_N(t)^\star$
(depending on the scatterers $\epsilon_j$) by $\Lambda_t^{\star}
\equiv \bigotimes \Gamma_{t}^{\star}$ (only depending on $\mu$)
and expectations at time $t$ become
\[
\Tr \big[ \Lambda_t\big(A_N \big) \, \,  \rho^N_0  \big]
\]
Computations are here even
simpler.\\

The results of section \ref{results} say essentially that
 the above effective dissipative dynamics reproduces the correct result
(see in fact Lemma  \ref{lemma:semigroup}).  For example,
\begin{equation}
\elim{N \uparrow +\infty} \frac{1}{N} \Tr\big[ M_N^z \, \rho^N_t
\big] = \lim_{N \uparrow +\infty} \frac{1}{N}
\Tr\big[\Lambda_t\big( M_N^z \big) \,\, \rho^N_0  \big] = \Tr
\big[ \Gamma_t(\sigma_{z}) \nu_0\big]
\end{equation}
In other words, the unitary dynamics \eqref{eq:QKM-UnitarySystem}
(depending on the scatterers $\epsilon_j$) is typically equivalent
with the effective dynamics $\Gamma_t$ on the one-particle level.
The marginal $\nu_t$ of \eqref{eq:reduc} exists and can be
obtained from $\nu_{t+1} = \Gamma^{\star}\nu_{t}$.  That specifies
the equation \eqref{mrel} and determines
\eqref{eq:QuaEntDensity}.\\

Note finally that the reduction of the unitary time evolution
\eqref{eq:QKM-UnitarySystem} to an effective dynamics on a
one-particle system remains non-Abelian. Of course, our
macroscopic observables start to commute,
\begin{displaymath}
\big[ \frac{M_N^x}{N} , \frac{M_N^y}{N} \big] = \mathcal{O}\bigg(
\frac{1}{N} \bigg)
\end{displaymath}
but $\Gamma^\star$ remains defined on density matrices and does
not yield a classical dynamics on diagonal elements of the density
matrix $\nu_0$.\\

As stated in Theorem \ref{result:relaxation}, as in the original
Kac ring model (or as for the rigorous derivation of the Boltzmann
equation), one really does not need this assumption of molecular
chaos to get relaxation.  It is replaced with statistical
assumptions on the level of the initial conditions for the
scatterers --- they are typical with respect to the probability
\eqref{eq:QKM-ProbCanonical} --- and for the spins
--- they are e.g. randomly sampled with a fixed magnetization in
\eqref{eq:ainidens}. In that case we think of both the $\eta$
(wavefunction) \`and the $\epsilon$ (classical scattering centers)
as dynamical variables.

\subsection{Autonomous equations and $H-$theorem. }
\label{subsec:auto}

The equation \eqref{mrel} is autonomous in the sense that the
value of $\vec{m}_t$ determines $\vec{m}_{t+1}$ once we know
$\vec{h}\equiv h\,\vec{n}= (e_+-e_-)\,\vec{n}/2$ in the
Hamiltonian \eqref{eq:QK12_Hamiltonian} and the fraction $\mu$ of
scatterers:
\begin{equation}\label{mrelax}
\vec{m}_{t+1} = \vec{m}_t -2\mu[(\vec{n}\times \vec{m}_t)\sin h
\cos h - \vec{n}\times(\vec{n} \times \vec{m}_t) \sin^2 h]
\end{equation}
A more suggestive expression is obtained by decomposing $\vec{m}$
into a parallel and perpendicular component along $\vec{n}$:
\[
\vec{m} = (\vec{m}\cdot \vec{n})\vec n +
\vec{n}\times(\vec{m}\times\vec{n}) \equiv \vec{m}_\parallel +
\vec{m}_\perp
\]
for which
\begin{equation}\label{deco}
\vec{m}_\parallel(t+1)=\vec{m}_\parallel(t),\quad
\vec{m}_\perp(t+1) = (1-2\mu\sin^2 h) \vec{m}_\perp(t) + 2\mu \sin
h \cos h \;\vec{m}_\perp(t)\times \vec{n}
\end{equation}
 The
relaxation formula \eqref{mrel} simplifies for certain initial
conditions. For example, when taking \eqref{eq:ainidens} or
\eqref{grand}, so that $m_0^x=m_0^y=0$, we get the damped
oscillator
\[
 m_t^z = m_e^z + (m_0^z - m_e^z) r^t \cos \omega t
\]
with $r^2\equiv (1-\mu)^2 + \mu^2 + 2 \mu(1-\mu) \cos 2h$ and
$\tan (\omega) \equiv \mu\sin 2h/(1-\mu +\mu\cos 2h)$. The
frequency is maximal for $h = \pi/4$ and the damping is maximal
for $h=\pi/2$. In that case, the relaxation of $m^z_t$ also looks
autonomous but that is because of the special initial data.\\
In general, $m^z_t$ does not relax autonomously and in contrast
with \eqref{result:entropy}, its associated entropy
\[
-\frac{1+m_t^z}{2} \log \frac{1 + m_t^z}{2} -
\frac{1-m_t^z}{2}\log \frac{1-m_t^z}{2}
\]
need not be monotone (even in the case of \eqref{eq:ainidens} or
\eqref{grand}).  Going to the bigger picture with three
magnetization components reveals a new structure in which the
entropy does increase. From \eqref{deco},we see that
$\|\vec{m}(t)\|$ decreases as
\begin{equation}\label{wesee}
\|\vec{m}_\perp(t+1)\|^2 = (1-4\mu(1-\mu)\,\sin^2
h)\,\|\vec{m}_\perp(t)\|^2
\end{equation}
and hence, combined with \eqref{result:entropy}, the monotonicity
 $s(t+1) \geq s(t)$ follows and strictly so if $\mu\in (0,1)$ and
$h\neq k\pi$. We thus see here that also in the quantum case,
autonomy on some macroscopic scale is intrinsically connected to
monotonicity of the corresponding entropy.

\subsection{Relaxation to equilibrium?}
\label{subsec:norelax}

The relaxation can be read off clearly from \eqref{deco}: we get a
spiral motion in the plane perpendicular to $\vec{n}$. The model
does however not show the full glory of relaxation to equilibrium.
That is already true in the original Kac ring model. As an
example, consider a macroscopic variable which involves a two-spin
function, in contrast with the single site observables that we
introduced in \eqref{eq:sma}:
\begin{displaymath}
A_N = \sum_j \sigma_{z,j}\sigma_{z,j+1}
\end{displaymath}
It is easy to verify that $A_N/N$ does not show relaxation.  This
is not surprising given the fact that two neighboring spins live
exactly the same history and thus they do not decorrelate. This
seriously restricts the usefulness of the model for studies of
relaxation but it remains possible to study the phenomenon of
relaxation for the special macroscopic observables of
\eqref{eq:sma}.\\

On the other hand, the fact that we consider relaxation in terms
of macroscopic observables should not be considered as a
restriction.  After all, equilibrium is characterized by a maximal
entropy condition given macroscopic constraints.  In particular,
relaxation is not be read off from the Liouville-von Neumann
evolution for all microscopic details of the density matrix.  The
fact that we first let $N\uparrow +\infty$ (before time) avoids
the presence of (quasi-)periodicities or of Poincar\'e-recurrence.

A final useful comparison concerns the notion of ergodicity which
is quite universal, see e.g. \cite{Frigerio,Evans}.  Consider a
quantum mechanical system with a discrete non-degenerate energy
spectrum $(E_n)$ with $(\phi_n)$ a complete set of orthonormal
energy eigenfunctions. The wavefunction at time $t$ is denoted by
$\psi(x,t) =\exp[-iHt]\psi (x)$ for initial $\psi(x)= \sum_n c_n
\phi_n(x)$.  The time-average of the expectation value of a
Hermitian $A$ is
\begin{displaymath}
\bar{A} \equiv \lim_{T\uparrow+\infty} \frac 1{T} \int_0^Tdt\int dx
\,\psi^\star(x,t) A\psi(x,t)
\end{displaymath}
It is rather easy to see in that case (but it remains true in a
much broader context) that
\begin{equation}
\label{eq:ergo} \bar{A} = \sum_n |c_n|^2 \int dx \,
\phi_n^\star(x) A\phi_n(x)
\end{equation}
which could be argued to correspond to the microcanonical
ensemble.  Note however that no use was made in the above of the
fact that $N$ is large nor of the fact that we consider
macroscopic variables.  The result \eqref{eq:ergo} is equally not
very satisfactory: we are interested in the typical manifest
behavior over realistic time-spans for large systems while
\eqref{eq:ergo} gives trivial information about an infinite
time-average that cannot be identified with relaxation phenomena.
There is in fact no relaxation to equilibrium on a microscopic
scale even though \eqref{eq:ergo} always holds.

\subsection{Higher spins}

Instead of considering spin 1/2 particles, it is also possible to
set up exactly the same problem as above for a Hilbertspace
$\mathcal{H}_N = \bigotimes_{j=1}^N \mathbbm{C}_j^n$.  We are then
dealing with $n\times n$ matrices but there is little difference
in the computations except for one extra complication: there is a
larger class of conserved quantities.  To explain that, we split
the single-site observable $A$ (a Hermitian matrix in
$\mathbbm{C}^{n \times n}$) in a part that commutes with the
Hamiltonian $H$, and its orthogonal complement:
\begin{equation}
\label{eq:splitting}
A = A_H \oplus A_{\bot}
\end{equation}
with $[A_H,H]=0$ and with respect to the scalar product
\begin{equation}
\label{eq:QKRes_DefinitionScalarProd}
\braket{A|B} = \Tr\big[´A^{\star} B \big]
\end{equation}
The set of commuting observables $\mathcal{A}_H$ is a vector space
with dimension
\begin{displaymath}
\textrm{dim}\big(\mathcal{A}_H\big) = n
\end{displaymath}
If $n>2$ there is more than just the $\mathbbm{1}$ and $H \in
\mathcal{A}_H$ (e.g. not all functions of $H$ can be expressed as
linear combinations of $\mathbbm{1}$ and $H$).\\
Because of the triviality of the dynamics as was discussed in the
previous section, these extra conserved quantities give rise to
conserved quantities for the full system dynamics and none of them
disappear as the size of the system grows. For more realistic
systems, truly interacting, we expect that, in the thermodynamic
limit, essentially only such physical quantities as particle
number and total energy remain conserved. In that respect, the
case $n=2$ is more physical and that is the reason why we have
concentrated on it since the beginning.  The mathematical theory
can however be completed if we change the definition of the
microcanonical ensemble \eqref{eq:fmicro} to include the extra
conserved quantities.

\section{Proof of results}
\label{sec:proofs}

Since it is important for physical interpretation that the
relaxation curve \eqref{mrel} is obtained for almost all choices
of scatterers $\epsilon$ drawn from \eqref{eq:epsset}, we must
first remind in Lemma \ref{lemma:equivalence of ensembles} of the
thermodynamic equivalence with a grand-canonical set-up.  Next, in
Lemma \ref{lemma:semigroup} is shown that an effective dynamics
reproduces the time-evolution of our macroscopic observables.
Using that, and from Lemma \ref{lemma:large deviations}, the
propagation of condition \eqref{eq:initialansatz} is obtained. It
is therefore sufficient to derive \eqref{mrel} to get
\eqref{eq:rel}.  That computation is introduced by Lemma
\ref{lemma:epslimmagnetization}.  The entropy
\eqref{result:entropy} and its monotonicity can then also be
computed using the obtained law of large numbers. \\

Let $\Omega \equiv \{0,1 \}$ and $\epsilon \in \Omega^{N}$. We
have already defined the ``canonical'' measure
$\mathbbm{P}^N_{\mu}$ in \eqref{eq:QKM-ProbCanonical} on
$\Omega_{\mu}^N$ of \eqref{eq:epsset}. Define now its
``grand-canonical'' version, the probability measure
$\mathbbm{P}_{\mu}$ on $\Omega^{\mathbbm N}$, as the Bernoulli
measure with density $\mathbbm{P}_{\mu}(\epsilon_j = 1)=\mu$.
$\mathbbm{E}^{N}_{\mu}(\cdot)$ and $\mathbbm{E}_{\mu}(\cdot)$
denote the expectation values with respect to
$\mathbbm{P}^{N}_{\mu}$ and $\mathbbm{P}_{\mu}$ respectively.  The
following equivalence is standard.\\

\begin{lemma}\label{lemma:equivalence of ensembles}
\!\!\!\! For all functions
$g \equiv g(\epsilon_1, \ldots, \epsilon_k)$ on $\Omega^{k}$
\begin{displaymath}
\Big| \, \mathbbm{E}^N_{\mu}(g)-\mathbbm{E}_{\mu}(g)\Big|
\leqslant \frac{2^{2k+1}k}{N}  \| g \| \qquad \textrm{if} \quad 2k
\leqslant N
\end{displaymath}
with
\begin{displaymath}
\| g \| \equiv \max_{\xi \in \Omega^{k}} |g(\xi)|
\end{displaymath}
\end{lemma}
\begin{proof}
\begin{displaymath}
\bigg| \mathbbm{E}^N_{\mu}(g) - \mathbbm{E}_{\mu}(g) \bigg|
\leqslant \sum_{\xi \in \Omega^{k}} | g(\xi)| \, \big|
\mathbbm{P}^N_{\mu}(\xi)-\mathbbm{P}_{\mu}(\xi) \big|
\end{displaymath}
The right-hand side can be bounded by developing
$\mathbbm{P}^N_{\mu}(\xi)$ in conditional expectations. By
explicit calculation
\begin{displaymath}
\big| \mathbbm{P}^N_{\mu}\big(\xi_{k+1}=1\mid
(\xi_{1},\ldots,\xi_{k})= \xi) \big)-\mu \big| \leqslant
\frac{k+1}{N-k}
\end{displaymath}
after which a simple calculation finishes the proof. ~\\~\\
\end{proof}

We define for fixed $j$ and matrix element per element
\begin{displaymath}
\Gamma_{t}(A_j) \equiv \mathbbm{E}_\mu A_{j}(t,\epsilon)
 =
\elim{N \uparrow +\infty}\mathbbm{E}_{\mu}^N A_{j}(t,\epsilon)
\end{displaymath}
where the last equality follows from the previous Lemma
\ref{lemma:equivalence of ensembles}.\\

\begin{lemma}
\label{lemma:semigroup} \!\!\!For macroscopic observables $A_N =
\sum_{j}A_{j}$ and continuous functions $f$, the dynamics is
typically equivalent to the semigroup dynamics $\Lambda_t$,i.e.,
\begin{displaymath}
\elim{N \uparrow +\infty} \Tr \Bigg[ f \bigg( \frac{A_N}{N} \bigg) \rho_{t}^N  \Bigg]
=
\lim_{N \uparrow +\infty} \Tr \Bigg[ f \Bigg( \Lambda_t
 \bigg( \frac{A_N}{N} \bigg) \Bigg) \rho_{0}^N  \Bigg]
\end{displaymath}
\end{lemma}
\begin{proof}
The proof is for all monomials. By the Stone-Weierstrass theorem,
this proves the theorem for all continuous functions $f$.\\
 Let
\begin{displaymath}
D(n)\equiv \{1, \ldots, N\}^n
\end{displaymath}
\begin{displaymath}
T(n,t) \equiv \Big\{ K \in D(n) \big| \, \forall i,j \leqslant n :
i\neq j \quad \Rightarrow \quad  |K_i-K_j|>t \Big\}
\end{displaymath}
and
\begin{displaymath}
A(K) \equiv  A^{k(1)}_{j(1)}\otimes \ldots \otimes A^{k(l)}_{j(l)}
\end{displaymath}
where
\begin{displaymath}
K=
 \{ \underbrace{j(1), \ldots ,j(1)}_{k(1)} ,
  \ldots ,\underbrace{ j(l), \ldots,j(l) }_{k(l)} \}
\end{displaymath}
We also need the index set $J(K)\equiv \{j \, | \, \exists p : K_p=j \}$.

We begin by calculating the $\epsilon$-average. Denote
$A_N(t,\epsilon) \equiv U_N(t)^{\star} A_N U_N(t)$, and similarly
for the one-site observables $A_j$.
\begin{align*}
\Tr \Bigg[ \bigg( \frac{A_N}{N} (t,\epsilon) \bigg)^{n} \rho_{0}^N
\Bigg] = \frac{1}{N^{n}} & \sum_{w} \sum_{K \in T(n,t)} \prod_{j
\in J(K) }^n \Tr \big[ A_{j}(t,\epsilon) \chi_{j,w}^{N}\big]
\prod_{j\notin J(K)} \Tr \big[ \chi_{j,w}^{N} \big] \\
& + \frac{1}{N^n} \sum_{K \in D(n) \setminus T(n,t)} \Tr \Big[
 A(K) (t,\epsilon) \rho_{0}^N \Big]
\end{align*}
The last term can be dropped because
\begin{displaymath}
\bigg| \frac{1}{N^{n}}\sum_{K \in D(n) \setminus T(n,t)} \Tr\Big[
A(K) (t,\epsilon) \rho_{0}^N \Big]\bigg|
\leq\frac{1}{N^{n}}\sum_{K \in D(n) \setminus T(n,t)} \| A(K) \| =
\mathcal{O} \bigg(\frac{\|A\|^n}{N} \bigg)
\end{displaymath}
We apply Lemma \ref{lemma:equivalence of ensembles} with
$g_{K,w}(\epsilon) \equiv \prod_{j \in K}\Tr
\big[A_{j}(t,\epsilon) \chi_{j,w}^{N}\big]$ as function on
$\Omega^{tn}$.  For $K \in T(n,t)$,
\begin{displaymath}
\bigg| \mathbbm{E}_\mu^N \prod_{j \in K} \Tr
\big[A_{j}(t,\epsilon) \chi_{j,w}^N\big]
 - \Big( \mathbbm{E}^{C}_\mu \Tr\big[A_{j}(t,\epsilon)
  \chi_{j,w}^N \big] \Big)^{n} \bigg|
\leqslant \frac{2^{2tn} 2(tn)}{N} C^n \| A \|^{n}
\end{displaymath}
We replace $\mathbbm{E}_\mu \Tr [A_{j}(t,\epsilon) \chi_{j,w}^N]$
by $ \Tr [\Gamma_t(A_{j}) \chi_{j,w}^{N}]$ and we add terms of
order $N^{-1}$ (essentially the ones we first subtracted) to
arrive at the average from the lemma.

For the variance, we similarly write
\begin{displaymath}
\mathbbm{E}^N_\mu \Bigg(\Tr \Bigg[ \bigg( \frac{A_N}{N}
(t,\epsilon) \bigg)^{n}  \rho_{0}^N \Bigg]\Bigg)^{2}
\end{displaymath}
as
\begin{displaymath}
\frac 1{N^{2n}}\sum_{K,L \in T(n,t)} \mathbbm{E}^N_\mu(  \Tr
\big[A(K)(t,\epsilon)\rho_{0}^N \big] \Tr \big[ A(L)(t,\epsilon)
\rho_{0}^N \big]) + \mathcal{O}\big(N^{-1}\big)
\end{displaymath}
Also the contribution of sets $K$ and $L$ that are within a
distance less than $t$ are of lower order. We apply
Lemma \ref{lemma:equivalence of ensembles} just as above.\\~\\
\end{proof}

\begin{lemma}
\label{lemma:large deviations} \!\!\!\! For all continuous
functions $f$ and macroscopic  observables $A_N$,
\begin{displaymath}
\elim{N \uparrow +\infty} \Tr \Bigg[ f \bigg( \frac{A_N}{N} \bigg)
\rho_{t}^N  \Bigg] = \lim_{N \uparrow +\infty} f \Bigg( \Tr\bigg[
\Lambda_{t}\bigg(\frac{A_N}{N}\bigg) \rho_{0}^N \bigg] \Bigg)
\end{displaymath}
at all times $t$.
\end{lemma}

\begin{proof}
By Lemma \ref{lemma:semigroup} we can insert the initial density
matrix.  The statement then follows from applying condition
\eqref{eq:initialansatz} to the observable $\sum_j \Gamma_t(A_j)$.\\~\\
\end{proof}

As an illustration of the lemma above, we show that it can be applied for the state
$\rho^N_0$ given by \eqref{eq:ainidens}.

\begin{example}\label{ex1}
Call $P_{a}^N$ the projector on the eigenvalue $a$ of the
macroscopic variable  $A_N/N$. Let $I^{N}$ be the spectrum of
$A_N/N$ and $I=\overline{\cup I^{N}}$.  We want to establish a
large deviations property for the probability measures $W^{N}$:
\begin{equation}\label{ww}
W^{N}(K)\equiv \sum_{a \in I^{N} \cap K}  \Tr
\big[P_{a}^{N}\rho_{0}^{N}\big]
\end{equation}
The product $P_{a}^{N}\rho_{0}^{N}$ can be written as a sum over
1-dimensional projections and \eqref{ww} can be developed and
written in terms of
\[ \Tr [P_{a_j} Q_{m_j}] \] which takes four possible values depending
 on $(p_j,m_j)$.
The problem amounts therefore to estimate the probability that at
the same time $\sum_j p_j, \sum_j m_j$ and $\sum_j m_j p_j$ take
on specific values when the $(p_j,m_j)$ are independent and
identically distributed with uniform weights.  That can be found
from the multinomial distribution and Stirling's formula gives the
large deviation rate function corresponding to \eqref{ww}. As a
consequence, a law of large numbers is satisfied and in the
thermodynamic limit the state \eqref{eq:ainidens} concentrates on
only one eigenvalue
of $A_N / N$.\\~\\
\end{example}

We now come back to \eqref{eq:reduc} defining the marginal
$\nu_0$. We split the Pauli matrices $\sigma_{\alpha} =
\sigma^{\alpha}_{H} + \sigma^{\alpha}_{\perp}$ as in
\eqref{eq:splitting} and parameterize in the single-site
Hamiltonian basis:
\begin{equation}
\label{eq:nu0inHbasis} \nu_0 = \begin{pmatrix} p & \beta \\
\bar{\beta} & 1-p
\end{pmatrix} \qquad
\sigma^{\alpha}_{\perp}=\begin{pmatrix} 0 & \theta_{\alpha} \\
\bar{\theta}_{\alpha} & 0 \end{pmatrix}
\end{equation}
The typical value of the magnetization is defined to be:
\begin{displaymath}
\vec{m}_t \equiv \elim{N \uparrow +\infty} \Tr \Bigg[
\frac{\vec{M}_N}{N} \rho_t^N  \Bigg]
\end{displaymath}
\begin{lemma}
\label{lemma:epslimmagnetization} \!\!\!\!
\begin{equation}\label{eps1}
\vec{m}_t = \vec{m}_e + \RE\Bigg[ \bigg(\vec{m}_0 - \vec{m}_e +
\frac{i(\vec{h}\times\vec{m}_0)}{h} \bigg) \Big(1-\mu+\mu
\exp[i(e_+ - e_-)]\Big)^t\Bigg]
\end{equation}
\end{lemma}
\begin{proof}
Fixing $t$, we write $k(\epsilon)\equiv \sum_{j=1}^t \epsilon_j$
and we obtain
\begin{displaymath}
\Gamma_{t}(\sigma^{\alpha}_{\perp}) = \mathbbm {E}_{\mu} \Big[
{U^{k(\epsilon)}}^{\star} \sigma^{\alpha}_{\perp}
{U^{k(\epsilon)}} \Big] =
\begin{pmatrix} 0 & z_{\alpha}(t) \\ \bar{z}_{\alpha}(t) & 0 \end{pmatrix}
\end{displaymath}
with
\begin{equation}\label{aha}
z_{\alpha}(t) \equiv \theta_{\alpha} \big(1 - \mu + \mu \exp[i(e_+
- e_-)]\big)^t
\end{equation}
$\sigma^{\alpha}_{H}$ is invariant under the time-evolution and it
is easy to verify that
\begin{displaymath}
\lim_{t \uparrow +\infty} \Tr \big[ \Gamma_t(\sigma_{\alpha})
\nu_0 \big]
 = \Tr\big[\sigma^{\alpha}_H \nu_0 \big]
 = m^{\alpha}_e
\end{displaymath}
This yields
\begin{displaymath}
m^{\alpha}_t = m^{\alpha}_e + \RE\Big[ 2 \bar{\beta}
\theta_{\alpha} \big(1 - \mu + \mu \exp[i(e_+ - e_-)]\big)^t \Big]
\end{displaymath}
Now it is a matter of writing $2 \bar{\beta} \theta_{\alpha}$ as
\begin{displaymath}
2 \bar{\beta} \theta_{\alpha} = \Tr[ \nu_0
\sigma^{\alpha}_{\perp}] - \frac{\Tr[H \nu_0
\sigma^{\alpha}_{\perp}]}{h}
\end{displaymath}
then inserting $H = \vec{h} \cdot \vec{\sigma}$, and recognizing
the resulting complex numbers as components of the vector product
$\vec{h} \times \vec{m_0}$, to find \eqref{eps1}
almost immediately.~\\~\\~\\
\end{proof}
The asymptotics \eqref{eq:QK12_Relaxation} is trivial and the
proof of Theorem \ref{result:relaxation} is thus completed if we
add the equation \eqref{result:entropy} for the entropy $
\mathcal{S}[\nu_t]$.

\begin{proof}
The explicit formula \eqref{result:entropy} can immediately be
obtained by using the Bloch representation:
\begin{displaymath}
\nu_t = \frac{\mathbbm{1}+\vec{m}_t \cdot \vec{\sigma}}{2}
\end{displaymath}
Inserting it in
the entropy density \eqref{eq:QuaEntDensity} yields the result.\\

Concerning the increase of entropy: denote by $\nu_e$ the one-site
marginal of the equilibrium state \eqref{pe}, with magnetization
$\vec{m}_e$. A short computation gives
\begin{displaymath}
\nu_e = \begin{pmatrix} p & 0 \\ 0 & 1-p \end{pmatrix}
\end{displaymath}
where $p$ is the first diagonal element of $\nu_0$, see
\eqref{eq:nu0inHbasis}.

The relative entropy between two marginals is defined as
\begin{equation}
\label{relentropy} \mathcal{S}[ \nu | \nu_e ] =  \Tr[ \nu \log \nu
] - \Tr[ \nu \log \nu_e]
\end{equation}
We immediately have:
\begin{displaymath}
\mathcal{S}[ \nu_t | \nu_e ] = \mathcal{S}[\nu_e] -
\mathcal{S}[\nu_t]
\end{displaymath}
since the $\Gamma$-dynamics works only on the off-diagonal
elements of $\nu_0$. Together with the entropy-contraction
inequality for completely positive maps
\begin{displaymath}
\mathcal{S}\big[\Gamma^{\star}(\nu_t) \big| \Gamma^{\star}(\nu_e)
\big] \leqslant \mathcal{S}[\nu_t|\nu_e]
\end{displaymath}
and since $\Gamma^{\star}(\nu_e) = \nu_e$, we find:
\begin{displaymath}
 \mathcal{S}\big[ \Gamma^{\star} (\nu_t)
\big] = \mathcal{S}[ \nu_{t+1} ] \geqslant \mathcal{S}[\nu_t]
\end{displaymath}
\end{proof}

\section*{Acknowledgments}
We are grateful to I. Callens, A. van Enter, B. Haegeman and S.
Pirogov for useful discussions.


\end{document}